\documentclass[%
 reprint,
 amsmath,amssymb,
 aps,
]{revtex4-2}

\usepackage{graphicx}
\usepackage{dcolumn}
\usepackage{bm}
\usepackage{xcolor}
\usepackage{dsfont}
\usepackage{pgfplots}
\pgfplotsset{compat=1.15}
\usepackage{mathrsfs}
\usetikzlibrary{arrows}

\usepackage[utf8]{inputenc}
\usepackage[T1]{fontenc}
\usepackage[english,serbian]{babel}
\usepackage[OT1]{fontenc}
\usepackage{mathrsfs}
\usepackage{graphicx}
\usepackage{amssymb}
\usepackage{amsbsy}
\usepackage{latexsym}
\usepackage{mathtools}
\usepackage{titlesec}
\usepackage{amsmath}
\usepackage{color}
\usepackage[utf8]{inputenc}
\usepackage{lipsum}
\usepackage{fontenc}
\usepackage{bbold}
\usepackage{tikz-cd}

\usepackage{chngcntr}
\numberwithin{equation}{section}
\renewcommand{\theequation}{\arabic{section}.\arabic{equation}}

\def\dj{d\kern-0.4em\char"16\kern-0.1em}

\def\diff{\textrm{d}}
\def\Diff{\textrm{D}}

\def\tr{{\rm Tr}}

\def\dj{d\kern-0.4em\char"16\kern-0.1em}
\def \Dj {\mbox{\raise0.3ex\hbox{-}\kern-0.4em D}}

\begin{document}
\definecolor{ududff}{rgb}{0.30196078431372547,0.30196078431372547,1.}
\definecolor{bcduew}{rgb}{0.7372549019607844,0.8313725490196079,0.9019607843137255}
\preprint{APS/123-QED}

\title{Boundary terms, branes and AdS/BCFT in first-order gravity}

\author{Du\v{s}an \Dj or\dj evi\'{c} and Dragoljub Go\v{c}anin}
\affiliation{Faculty of Physics, University of Belgrade, Studentski Trg 12-16, 11000 Belgrade, Serbia}

\email{djdusan@ipb.ac.rs}

\selectlanguage{english}


\begin{abstract}

We provide an account of the issue of Gibbons-Hawking-York-like boundary terms for a gravity theory defined on a Riemann-Cartan spacetime. We further discuss different criteria for introducing boundary terms in some familiar first-order gravity theories with both on-shell vanishing and non-vanishing torsion, along with considerations regarding the thermodynamics of black holes and profiles of the End-of-the-World branes. Our analysis confirms the expected geodesic profile of the End-of-the-World brane in the BF formulation of Jackiw-Teitelboim gravity. Finally, we present the first realisation of the AdS/BCFT duality for spacetime with torsion.

\end{abstract}

\maketitle

\selectlanguage{english}

\section{Introduction}
\label{introduction}

It is well-known that the four-dimensional (4D) Einstein-Hilbert (EH) action of General Relativity (GR),
\begin{equation}\label{EHg}\frac{1}{16\pi G}\int \mathrm{d}^4x \sqrt{-g}R,
\end{equation}
involves second-order derivatives of the metric $g_{\mu\nu}$ and thus requires an additional boundary term in order to have a well-posed variational problem \cite{Wald:1984rg}. This boundary term is called the Gibbons-Hawking-York (GHY) term, 
\begin{equation}\label{ghy}
    \frac{1}{8\pi G}\int \mathrm{d}^3y\; \epsilon \sqrt{|h|}K,
\end{equation}
where $K$ is the trace of the extrinsic curvature of the boundary,  $h_{\alpha\beta}$ is the induced metric on the boundary, while $\epsilon=\pm 1$ depends on the signature of the boundary. The GHY term
exactly cancels the boundary term in the variation of (\ref{EHg}) that involves $\partial\delta g_{\mu\nu}$. On the other hand, in the first-order formulation, the EH action is given by 
\begin{equation}\label{prvi}
\frac{1}{32\pi G}\int \varepsilon_{abcd}R^{ab}e^ce^d, 
\end{equation}
with $R^{ab}=\diff\omega^{ab}+\omega^{a}_{\;\;c}\omega^{cb}$ and clearly does not contain second derivatives of any fundamental field ($e^{a}$ and $\omega^{ab}$). Therefore, we cannot hold to the same argument as before for adding a boundary term to the action (\ref{prvi}). 

Local analysis of equations of motion is usually done without even thinking about boundary conditions. In this context, one could simply ignore the question altogether. However, boundary terms play an important role in the study of black hole thermodynamics. Namely, one way to obtain the celebrated Bekenstein-Hawking formula for the entropy of a Schwarzchild black hole is to compute the Eucledian path integral \cite{PhysRevD.15.2752}. As $R=0$ for a vacuum solution, the bulk action is equal to zero on-shell. Therefore, the total contribution comes entirely from the (properly regularized) GHY boundary term (\ref{ghy}), which in turn reproduces the desired area-entropy law.

The importance of boundary terms in the Hamiltonian analysis of gravity is also well-known since the work of Regge and Teitelboim \cite{REGGE1974286}. 
However, what we will be concerned with in this paper is the study of boundary terms in the context of AdS/CFT duality, initially introduced in the string theory setup \cite{Maldacena:1997re}. Besides general considerations regarding asymptotic boundary conditions based on holographic arguments, we focus on the role that boundary terms have in establishing the AdS/BCFT correspondence 
\cite{Takayanagi:2011zk, Fujita:2011fp}. As the holographic duality relates a gravity theory living on an asymptotically anti-de Sitter (AdS) spacetime $\mathcal{N}$ and the corresponding conformal field theory (CFT) living on the asymptotic boundary $\mathcal{M}=\partial\mathcal{N}$, the CFT dual naturally has no boundary ($\partial^2\mathcal{N}=\emptyset$). Nevertheless, there are cases (e.g. open string world-sheet theory) where the CFT boundary is present. To include those situations, one is led to consider the so-called boundary CFT (BCFT). Those are CFTs with a boundary and suitable boundary conditions. To apply the holographic ideas in this scenario, we have to introduce a brane $\mathcal{Q}$, sharing a common boundary with the CFT, such that the boundary of the bulk $\mathcal{N}$ is given by $\partial\mathcal{N}=\mathcal{M}\cup \mathcal{Q}$ and  $\partial\mathcal{M}=\partial\mathcal{Q}$. This brane is called the End-of-the-World (EOW) brane. EOW branes were previously studied in the language of first-order gravity in \cite{Takayanagi:2020njm}, where the first-order gravity action was accompanied by the GHY term. We will take the bottom-up approach \cite{Harvey:2023pdv}, where we have to model our action. 
Since the GHY term has many different motivations, even in Einstein's gravity \cite{Dyer:2008hb}, we first have to discuss various aspects of GHY-like boundary terms in first-order gravity to ensure that their inclusion in the action is indeed well-justified. 

The main motivation for our work comes from a realization that torsion can be significant in holographic duality, as it can help us introduce the boundary spin-current and thus analyse strongly coupled systems with spin transport \cite{Gallegos:2020otk, Erdmenger:2022nhz, Banados:2006fe, Cvetkovic:2017fxa}. With that in mind, it is natural to consider a situation where the dual field theory has a boundary and, for that matter, we should generalize the AdS/BCFT correspondence to Riemann-Cartan bulk geometry. However, this seems a nontrivial task, and so we choose to tackle simple models of bulk gravity that involve torsion (similar logic was used in  \cite{Blagojevic:2003uc, Blagojevic:2003wn, Blagojevic:2004fu, Blagojevic:2003vn} to study consequences of Riemann-Cartan geometry on asymptotic symmetries).

The plan of the paper is the following. In the next section, we provide an account of various reasons for introducing (or not) boundary terms in first-order gravity. Based on this, in Section III, we analyse the case of $\textrm{2D}$ BF theory and confirm the expected profile of the EOW brane. In section IV, we consider the AdS/BCFT setup for $\textrm{3D}$ Chern-Simons gravity with torsion and determine how torsion modifies the profile of the EOW brane. We summarise our conclusions in Section V.       

\section{Boundary terms}

By going through the literature, one can find several different justifications for introducing gravitational boundary terms in first-order formalism. 
The choice of boundary terms is naturally related to a given set of boundary conditions. Perhaps the most obvious question one can ask is whether first-order gravity can be consistently formulated by fixing both the spin connection $\omega^{ab}$ and the vielbein $e^{a}$ at the boundary, i.e. by imposing Dirichlet boundary conditions on both fields? Note that there are situations when this is not possible. For example, in the case of three-dimensional gravity, the conjugate variable to the spin connection is precisely the vielbein field \cite{Witten:1988hc}. This means that we cannot consistently fix both fields at the boundary, just like we cannot fix both the coordinate and its conjugate momentum of a quantum particle (note that one should use Dirac brackets when dealing with constrained systems  \cite{Ferrari:1996wq}).
Also, a rather surprising fact that the GHY boundary term for the second-order formulation of EH gravity appears to be the right choice in so many different respects raises the question of whether there is a similar boundary term of the same importance in first-order gravity. Before answering these questions, we discuss the main rationales for introducing boundary terms in first-order gravity.    

\subsection{No boundary terms}

To begin with, let us again point out that the main reason for introducing the GHY boundary term in the metric formulation of EH gravity is precisely the second-order character of the EH action. It seems reasonable, therefore, to consider an option to simply not introduce any GHY-like term in first-order gravity. 
In that case, for example, four-dimensional gravity would be determined solely by EH action (\ref{prvi}). This, however, may seem unsatisfactory because the direct, naive, computation of the black hole entropy in first-order formalism using Euclidean path integral and on-shell action would yield zero.  
One could naively claim that the entropy of a black hole in the Einstein-Cartan theory of gravity should indeed be zero based on this calculation. However, this position could only be sustained if one regards the quantum theory of gravity based on the metric formulation, where we integrate over metric in the path integral, to be different from the path integral quantum theory of gravity in first-order formulation, schematically, 
\begin{equation}
    \int \mathcal{D}g\; e^{-S[g]}\neq \int \mathcal{D}e \mathcal{D}\omega e^{-S[e,\omega]},
\end{equation}
even though the classical equations of motion are the same. Nevertheless, this should not matter for the semi-classical calculation of black hole entropy, as there are many calculations of entropy in the Riemann-Cartan theory that yield a non-zero answer, for example, by using the Nester formula \cite{Nester:1991yd}. Note also that one has to be very careful when using the Euclidean path integral on spacetimes with torsion \cite{Blagojevic:2006jk}.

For that matter, let us introduce the Nester formula \cite{Nester:1991yd, Chen:2015vya}. It is a well-known fact that Killing vector fields give rise to conserved charges. Charges are given as integrals over spatial infinity of a certain form, defined solely from the bulk action and a particular solution (properly renormalized using background spacetime with no horizon). For the EH action (\ref{prvi}), the conserved charge (energy) is given by 
\begin{equation}
Q_{\xi}= G\int _{S^2} \frac{1}{2} (\iota_{\xi}\omega^{ab})\Delta \rho_{ab}+\frac{1}{2}\Delta\omega^{ab}(\iota_{\xi}\overline{\rho}_{ab}).
\end{equation}
Here, $\rho_{ab}=\frac{\partial \mathcal{L}}{\partial R^{ab}}=\frac{1}{16\pi G}\varepsilon_{abcd}e^{c}e^{d}$ is determined purely from the bulk action, and $\iota_{\xi}$ denotes a contraction with Killing vector field $\xi$. To get a finite answer, $\Delta \rho_{ab}$ is defined as a difference $\Delta\rho_{ab}=\rho_{ab}-\overline{\rho}_{ab}$, with $\overline{\rho}_{ab}$ being computed for a geometry without a horizon. For example, using the Schwarzschild solution with $\xi=\partial_t$,
Nester's formula yields the result $E=M$, which, together with the first law of black hole thermodynamics $T\diff S=\diff E$ gives a well-known result for the entropy of a Schwarzschild black hole. Note that, for asymptotically AdS black holes, one can use the AdS/CFT correspondence to calculate the energy and from it the entropy of a black hole \cite{Tetradis:2011jn}.

Not adding any boundary term when computing the black hole entropy using the Euclidean on-shell action is advocated in \cite{Cvetkovic:2017nkg}.

\subsection{Moving the variation to the vielbein}\label{2B}

To motivate this point of view, let us consider statistical physics. A partition function in the canonical ensemble is defined as the sum of Boltzmann factors over all possible states of the system,
\begin{equation}
Z=\sum_{\mathrm{s}}e^{-\beta E_s}.
\end{equation}
This ensemble is usually defined as the ensemble at a fixed temperature ($T=\beta^{-1}$), while no other potential is held fixed (charges, like the volume or the total number of particles, are assumed fixed). The grand canonical ensemble is, on the contrary, defined by assuming that the chemical potential is fixed while the corresponding charge - the total number of particles - can vary. Note that we can still formally work in the grand canonical ensemble even if the number of particles $N$ is fixed; one simply has to impose the constraint equation $\langle N\rangle=N$, which amounts to an implicit relation between the chemical potential and the temperature. This is very similar to the relation between black hole thermodynamics in the first versus second-order formalism. Euclidean path integral naturally fixes the temperature using periodicity in the Euclidean time. In this way, the temperature is determined solely from the metric (i.e. the vielbein). Therefore, fixing the vielbein at the boundary implies working with the canonical ensemble (see, for example, \cite{Rakic:2023vhv} for a similar discussion on rotating black holes in second-order gravity, or \cite{Davis:2004xi} for dilaton gravity). On the other hand, if we were to fix both the vielbein and the spin connection at the boundary, we should work in some more general ensemble. 
For example, in the case of first-order EH action (\ref{prvi}), the on-shell variation is just the boundary term \cite{Liko:2008rr, Ashtekar:2008jw,Corichi:2016zac}
\begin{equation}
\int\varepsilon_{abcd}\delta \omega^{ab}e^ce^d.
\end{equation}
To make a transition to the canonical ensemble requires moving the variation from $\omega$ to $e$; this is effected by adding to (\ref{prvi}) the following boundary term,
\begin{equation}\label{varbt}
-\int \varepsilon_{abcd} \omega^{ab}e^ce^d. 
\end{equation}
Note that this boundary term is not manifestly covariant under local Lorentz transformations; a way to make it covariant was discussed in \cite{Liko:2008rr}. Namely, one introduces a normal to the boundary $n^a$ and rewrites (\ref{varbt}) in a manifestly covariant form
\begin{equation}
2 \int  \varepsilon_{abcd} n^a\Diff n^b e^ce^d.
\end{equation} 
This gives GHY term.
Unfortunately, it is not always possible to perform this operation and thus cannot be used as a universal criterion. For example, we can easily check that the five-dimensional term 
\begin{equation}\label{tro}
\int \varepsilon_{abcde}R^{ab}R^{cd}e^e
\end{equation}
is such that its on-shell variation contains the boundary term
\begin{equation}
\int\varepsilon_{abcde}\delta \omega^{ab}\omega^c_{\;\;f}\omega^{fd}e^e, \end{equation}
where we cannot move the variation from $\omega$ to $e$ by adding a suitable boundary term to the action. Incidentally, note that the quadratic term (\ref{tro}) yields torsion undetermined on-shell, which means that the spin connection cannot be expressed in terms of the vielbein. This is similar in spirit to the fact that the total number of particles does not have to be fixed in the grand canonical ensemble. 


\subsection{Holographic considerations}\label{2C}

The holographic dictionary establishes a relation between an on-shell bulk action and the generating functional of the dual CFT. The leading-order terms in the asymptotic expansion of the bulk fields play the role of operator sources in the boundary theory and, therefore, should be associated with Dirichlet boundary conditions. On the other hand, fields that appear in the expectation values of the holographic currents are left free to vary \cite{Andrade:2006pg}. A generic first-order variation with AdS asymptotics does not lead to this set of boundary conditions. Therefore, one has to add suitable boundary terms. As those boundary terms are added for the sake of holographic boundary conditions, we refer to them as GHY-like boundary terms \cite{Gallegos:2020otk}. Let us demonstrate that for three-dimensional gravity, the on-shell value of the boundary term that we add in the holographic set-up coincides with the 
GHY term.

To analyse the behaviour of bulk fields near the asymptotic boundary $\partial\mathcal{M}$, we use the Fefferman-Graham (FG) expansion of the bulk fields organized in powers of the radial coordinate $\rho$; the asymptotic boundary is located at $\rho=0$. Boundary fields (written with a tilde) are finite and do not have $\mathrm{d}\rho$ component. The expansions relevant for us are given by \cite{Cvetkovic:2017fxa, Klemm:2007yu, Banados:2006fe}
\begin{align}\nonumber
   & e^i=\frac{1}{\sqrt{\rho}}\left(  \tilde{e}^i+\rho \tilde{k}^i\right), \hspace{4mm} \omega^{ij}=\tilde{\omega}^{ij},\\
   & \omega^{i1}=\frac{1}{\sqrt{\rho}}\left( \tilde{e}^i-\rho \tilde{k}^i\right),\label{fg}
\end{align}
where $i,j=0,2$ (non-radial components). First, note that in the FG gauge, the GHY term can be written as (analogous to (\ref{varbt}))
\begin{equation}
- \int_{\partial\mathcal{M}}\varepsilon_{abc}\omega^{ab}e^c.
\end{equation} 
Expanding the bulk fields using (\ref{fg}) and focusing only on the finite part (nonzero, non-divergent part as $\rho\rightarrow 0$), we can see that the finite part vanishes. On the other hand, using the procedure described in \cite{Klemm:2007yu}, one concludes that the holographic boundary conditions force us to add a boundary term 
\begin{equation}
4\int_{\partial\mathcal{M}}\varepsilon_{ij}\tilde{e}^i\tilde{k}^j.
\end{equation}
Using the on-shell identity $\mathrm{d}\tilde{\omega}^{ij}+2\tilde{e}^i\tilde{k}^j+2\tilde{k}^i\tilde{e}^j=0$, which comes from the bulk equations of motion, we see that the last integral vanishes, as it is proportional to $\int_{\partial\mathcal{M}}\varepsilon_{ij}\mathrm{d}\tilde{\omega}^{ij}$, which equals zero assuming $\partial^{2}\mathcal{M}=\emptyset$. 
\subsection{Fixing the induced fields}\label{fixing}
The most recent approach to the problem of boundary terms is presented in \cite{Erdmenger:2022nhz}. First, let us point out that in the case of second-order gravity with the GHY boundary term, one only needs to fix the induced metric on the boundary, rather than the bulk metric itself, to still have a well-defined variation problem. One could even insist that this is a fundamental feature of Dirichlet boundary conditions. 
Therefore, it would be necessary to make sure that the appropriate boundary terms are added in order to fix only the induced fields. It was shown in \cite{Erdmenger:2022nhz} that this results precisely in the standard GHY term in the case of first-order EH gravity. 

Consider another example. In the case of five-dimensional Lovelock gravity, discussed in \cite{Cvetkovic:2017nkg}, with action 
\begin{align}\nonumber
\int\varepsilon_{abcde} \Big(&\frac{\alpha_1}{3}R^{ab}e^ce^de^e+ \alpha_2 R^{ab}R^{cd}e^e \\
& +\frac{\alpha_0}{5}e^ae^be^ce^de^e\Big), 
\end{align}
the appropriate boundary term that ensures we fix only the induced fields is given by 
\begin{align}\label{cvele}\nonumber
-2\int\varepsilon_{abcd}\Big( &\frac{\epsilon\alpha_1}{3}K^ae^be^ce^d+2\epsilon\alpha_2K^aR^{bc}e^d\\
&-\frac{2\alpha_2}{3}K^aK^bK^ce^d  \Big).
\end{align}
One can compute the (Euclidean) on-shell value of this boundary term on a black hole geometry from \cite{Cvetkovic:2017nkg}. The metric for this solution is 
\begin{align}\
    \diff s^2=&-f^2(r)\diff t^2+\frac{1}{f^2(r)}\diff r^2 \\\nonumber
    &+r^2(\diff \psi^2+\sin^2\psi \diff \theta^2+\sin^2\psi\sin^2\theta \diff \varphi^2),
\end{align}
where $f^2(r)=\frac{\alpha_1}{8\alpha_2}\left(r^2-\frac{r^8_{+}}{r^6}  \right)$.
The relevant components of the vielbein and the spin connection can be found in \cite{Cvetkovic:2017nkg}. 
Inserting these expressions in the (Euclidean) boundary term (\ref{cvele}) yields 
\begin{align}
\frac{8\alpha_1\alpha_2r\beta}{\alpha_2 l^2}\int_{S^3}e^2e^3e^4+\frac{\alpha_1^2r\beta}{\alpha_2}\int_{S^3}e^2e^3e^4=0,\end{align}
where $e^2=r\diff \psi$, $e^3=r\sin \psi \diff \theta$, $e^4=r\sin \psi \sin \theta \diff \varphi$, and $\beta$ is the inverse temperature. 
This proves that even if we insist on adding the GHY term to the action in \cite{Cvetkovic:2017nkg}, the final answer for the entropy of a black hole would remain zero. In a way, this is a consistency check, as the same result is obtained using the Nester's formula.

\subsection{Amplitude composition}

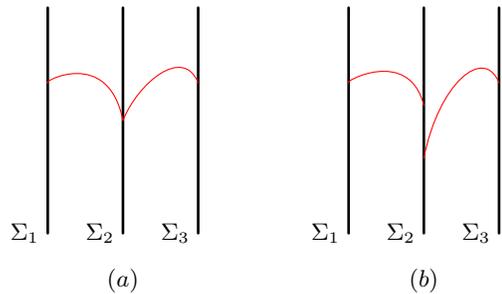
\begin{figure}[h!]
   \begin{tikzpicture}[line cap=round,line join=round,x=1.0cm,y=1.0cm]
\clip(-1.0,-0.9444861226016785) rectangle (7.0,3.22167500917438);
\draw [line width=1.pt] (0.,3.)-- (0.,0.);
\draw [line width=1.pt] (1.,3.)-- (1.,0.);
\draw [line width=1.pt] (2.,3.)-- (2.,0.);
\draw [color=red] (0.,2.) .. controls   (-0.1, 2.) and (0.8, 2.5) .. (1.,1.5);
\draw [color=red] (1.,1.5) .. controls   (1.2, 2.) and (1.8, 2.5) .. (2.,2.);
\draw [line width=1.pt] (4.,3.)-- (4.,0.);
\draw [line width=1.pt] (5.,3.)-- (5.,0.);
\draw [line width=1.pt] (6.,3.)-- (6.,0.);
\draw [color=red] (4.,2.) .. controls   (3.9, 2.) and (4.8, 2.5) .. (5.,1.7);
\draw [color=red] (5.,1.) .. controls   (5.2, 2.) and (5.8, 2.5) .. (6.,2.);
\draw (1.0,-0.9) node[anchor=south] {$(a)$};
\draw (5.0,-0.9) node[anchor=south] {$(b)$};
\draw (0.0,-0.0) node[anchor=east] {$\Sigma_1$};
\draw (1.0,-0.0) node[anchor=east] {$\Sigma_2$};
\draw (2.0,-0.0) node[anchor=east] {$\Sigma_3$};
\draw (4.0,-0.0) node[anchor=east] {$\Sigma_1$};
\draw (5.0,-0.0) node[anchor=east] {$\Sigma_2$};
\draw (6.0,-0.0) node[anchor=east] {$\Sigma_3$};
\end{tikzpicture}
\caption{Transition from hypersurface $\Sigma_1$ to $\Sigma_3$, with intermediate hypersurface $\Sigma_2$. In the second-order formalism, the first derivative in the direction normal to the boundary is discontinuous, and therefore the second derivative gives a delta function. In the first-order formulation, some components of the spin connection are discontinuous, and therefore its first derivative contains the delta function.}
\label{skica}
\end{figure}

Let us discuss one more important feature of the GHY term promoted in \cite{Hawking:1979ig}, in relation to the previous discussion. Assuming the validity of the path integral approach to quantum gravity, 
we can compute the transition amplitude from $\Sigma_1$ to $\Sigma_3$, in terms of induced fields at these hypersurfaces (Fig. \ref{skica}). We emphasize that the relevant fields are the induced ones, which is precisely the connection with the previous subsection. In the case of metric formulation of gravity, the insertion of a hypersurface $\Sigma_3$ would lead to a potential discontinuity of the metric derivatives in normal directions (schematically displayed as the red line on $(a)$, color online). As action contains the second-order derivative of the metric, a discontinuity of the derivative yields a Dirac delta function, and the GHY is added precisely to account for this additional term that would spoil the composition rule of the transition amplitudes. In the first-order formalism, some components of the spin connection do not have to be fixed on the hypersurface $\Sigma_{2}$, and its first derivative gives the delta function, thus also motivating the introduction of the GHY term.

Finally, it is important to note that there are cases where the three motivations for adding boundary terms we discussed so far: placing the variation only on the vielbein (discussed in subsection (\ref{2B})), holographic demands (discussed in subsection (\ref{2C})) and fixing only the induced fields on the boundary (discussed in subsection (\ref{fixing})), give different answers regarding the relevant boundary terms, even when all three of them are applicable. For example, let us briefly mention the case of four-dimensional gravity defined by the Holst action with the Barbero-Imirizzi parameter $\gamma$, often used in loop quantum gravity. The bulk action is given by \cite{Durka:2011yv}
\begin{align}\label{host}
\kappa\int\varepsilon_{abcd}&\left(R^{ab}e^ce^d+ \frac{1}{2}e^ae^be^ce^d\right)+\frac{2\kappa}{\gamma}\int R^{ab}e_ae_b.
\end{align}  
Now, the first method implies that the following boundary term should be added \cite{Corichi:2010ur, Corichi:2012zz}
\begin{equation}-\frac{2\kappa}{\gamma} \int\omega^{ab}e_{a}e_{b}.
\end{equation}
Making this term covariant \cite{Corichi:2016zac} yields 
\begin{equation}\label{tori}
-\frac{2\kappa}{\gamma} \int T^a(\delta^b_a-n^bn_a)e_b.
\end{equation}
On the other hand, the other two approaches suggest that there should be no boundary terms. One might say that this discrepancy is probably irrelevant as the boundary term (\ref{tori}) turns out to be zero on-shell, however, this only holds for pure gravity; by adding matter fields in the bulk, the issue regarding the choice of the boundary term rises again. This is not the only case where we find the discrepancy between methods from (\ref{2B}), (\ref{2C}) and (\ref{fixing}). Here, we also mention the case of \textrm{5D} CS gravity, which is of special interest for holography. 
The holographic methodology of (\ref{2C}) gives \cite{Gallegos:2020otk}
\begin{equation}\label{gao}
4\kappa\int \varepsilon_{ijkl}\left ( \tilde{R}^{ij}+ \tilde{e}^i \tilde{k}^j  \right)\tilde{k}^k \tilde{e}^l. 
\end{equation}
On the other hand, using the asymptotic expansions (\ref{fg}) (valid also for \textrm{5D} CS gravity) we find that the finite part of the boundary term (\ref{cvele}) is zero, a result that differs from (\ref{gao}) even on-shell. Of course, there might be another way to relate those two boundary terms, see \cite{Erdmenger:2022nhz}, but for now, this relation remains unclear. 
Moreover, as we noted earlier, the method presented in (\ref{2B}) is not even applicable in this case, due to the presence of the Gauss-Bonet term. Finally, there is the case of Mielke-Baekler (MB) gravity, which we will consider in section \ref{sectorsion}.

\section{EOW brane in 2D BF gravity}
Jackiw–Teitelboim (JT) gravity \cite{Jackiw:1984je, Teitelboim:1983ux,Iliesiu:2019xuh}
\begin{equation}
S_{JT}=\kappa\int\diff^{2}x\sqrt{-g}\Phi(R-\Lambda)+2\kappa \int\diff y\sqrt{|h|}\Phi K,  \end{equation}
is one of the most studied models of two-dimensional dilaton gravity because it offers the possibility to tackle the metric equations analytically and even work in the quantum regime. 
Moreover, the equations of JT gravity can be obtained from a topological BF gauge theory for a suitable gauge group. 
Boundary terms in the first-order JT gravity have been discussed in the past (see, for example
\cite{Grumiller:2017qao,Bergamin:2005pg,Grumiller:2020elf,Ebert:2022ehb}). 
We will consider the BF gauge theory to illustrate various approaches to boundary terms discussed in the previous section. 

The bulk ($\mathcal{N}$) BF action for gauge group $SO(2,2)$ is 
\begin{align}\label{BF1}
\kappa\int_{\mathcal{N}} \varepsilon_{\hat{A}\hat{B}\hat{C}}\phi^{\hat{A}} F^{\hat{B}\hat{C}},  
\end{align}
or in component form
\begin{align}\label{bf}
&\kappa\int_{\mathcal{N}}\varepsilon_{ab}\left[\varphi\left(R^{ab}+e^{a}e^{b}\right)+2\phi^{a}T^{b}\right],
\end{align}
where
\begin{align}
R^{ab}&=\diff\omega^{ab}+\omega^{a}_{\;\;c}\omega^{cb},\\
T^{a}&=\Diff_{\omega} e^{a}=\diff e^{a}+\omega^{a}_{\;\;b}e^{b},
\end{align}
are the curvature and torsion. Indices $a,b=0,1$ are ordinary Lorentz $SO(1,1)$ indices (see Appendix A for more details on the underlying algebraic structure of the theory). After partial integration, we get
\begin{align}\label{BF2}
\kappa\int_{\mathcal{N}}\varepsilon_{ab}\left[\varphi\left(R^{ab}+e^{a}e^{b}\right)-2\Diff\phi^{a}e^{b}\right]
+2\kappa\int_{\partial\mathcal{N}}\varepsilon_{ab}\phi^{a}e^{b}. 
\end{align}
The variation of action (\ref{BF2}) with respect to the spin connection comes down to
\begin{align}
\kappa\int_{\mathcal{N}}\left(
2\varepsilon_{ab}\phi_{c}e^{b}-\varepsilon_{ac}\diff\varphi\right)\delta\omega^{ac}+\kappa\int_{\partial\mathcal{N}}\varepsilon_{ab}\varphi\delta\omega^{ab}.
\end{align}
Based on the previous discussion in (\ref{2B}), if we want to transfer the variation from $\omega$ in the boundary term, we have to modify the BF action by including an additional boundary term, namely \cite{Bergamin:2007aw}
\begin{equation}\label{prvibound}
-\kappa \int_{\partial\mathcal{N}} \varepsilon_{ab}\varphi\omega^{ab}.
\end{equation}
Now, in the context of AdS/BCFT construction, the boundary $\partial\mathcal{N}$ consists of two parts: the asymptotic boundary $\mathcal{M}$ and the EOW brane $\mathcal{Q}$, that is $\partial\mathcal{N}=\mathcal{M}\cup\mathcal{Q}$. On the asymptotic boundary $\mathcal{M}$, we impose the usual Dirichlet boundary conditions that fix the boundary value of the bulk fields. On the other hand, we impose Neumann boundary conditions on the EOW brane $\mathcal{Q}$ that yield certain constraints on the boundary fields that dynamically determine the profile of the brane.

Next, we use reasoning from (\ref{2C}). The asymptotic FG expansion of the bulk fields $\omega^{01}$ and $\varphi$  that appear in the boundary term (\ref{prvibound}) is given by
\begin{align}
\label{omega}
&\omega^{01}=\frac{1}{\sqrt{\rho}}(\Tilde{e}-\rho \Tilde{k}), \\
\label{varfi}
&\varphi=\frac{1}{\sqrt{\rho}}(\Tilde{\varphi}+\rho\Tilde{\psi}).
\end{align} 
Let us prove that the finite part of the boundary term (\ref{prvibound}) is the same as the boundary GHY-like term that follows from the holographic consideration in \cite{Dordevic:2023crm}. Namely, for the asymptotic boundary $\mathcal{M}$, we have 
\begin{align}\nonumber 
-\int_{ \mathcal{M}}\varepsilon_{ab}\varphi \omega^{ab}&=-2\int_{ \mathcal{M}}\varphi\omega^{01}\\\nonumber&=-2\int_{\mathcal{M}} \frac{1}{\rho}(\tilde{\varphi}+\rho \tilde{\psi})(\tilde{e}-\rho \tilde{k}),
\end{align}
where in the final step we applied the FG expansion of the bulk fields. 
The finite part is therefore
\begin{align}
-2\int_{\mathcal{M}}(-\tilde{\varphi} \tilde{k} +\tilde{\psi} \tilde{e})&=4\int_{ \mathcal{M}}\tilde{\varphi}\tilde{k},
\end{align}
where in the last equality, we applied the equations of motion and removed the boundary term as before (since \cite{Dordevic:2023crm} deals with the standard situation where the only boundary of the bulk spacetime is the asymptotic boundary where the dual field theory lives). This further supports the validity of our boundary term (\ref{prvibound}).

We further focus on the EOW brane $\mathcal{Q}$ and make the boundary term (\ref{prvibound}) manifestly covariant by introducing a normalized vector field $n^{a}$ on $\mathcal{Q}$, yielding 
\begin{align}\nonumber
2\kappa \int_\mathcal{Q} \varepsilon_{ab}\varphi n^a \mathrm{D}n^b.
\end{align}
Note that we do not assume that $n^{a}$ is fixed off-shell, and in particular, that it is orthogonal to $\mathcal{Q}$; this can be achieved dynamically (on-shell) by inserting the projector $P^a_b=\delta^a_b-n^a n_b$ to the EOW brane $\mathcal{Q}$ in the boundary term from (\ref{BF2}). The total action becomes 
\begin{align}\nonumber
&\kappa\int_{\mathcal{N}}\varepsilon_{ab}\left [\varphi\left(R^{ab}+e^a e^b\right)- 2\mathrm{D}\phi^a e^b\right ]\nonumber\\
+2&\kappa\int_\mathcal{Q} \varepsilon_{ab}\phi^a(\delta^b_c-n^b n_c)e^c
+2\kappa \int_\mathcal{Q} \varepsilon_{ab}\varphi n^a \mathrm{D}n^b\nonumber\\
+2&\kappa T\int_\mathcal{Q}\varepsilon_{ab}n^a e^b,
\end{align}
where we also included the brane tension term ($T$ being the tension). 

Varying this action with respect to the fundamental fields and imposing Neumann boundary conditions on $\mathcal{Q}$, one finds the following. The $\delta\omega$ variation yields
\begin{equation}
\kappa \int_\mathcal{Q} \varepsilon_{ab}\varphi\delta \omega^{ab}+2\kappa \int_\mathcal{Q} \varepsilon_{ab}\varphi n^a \delta \omega^b_{\;\;c} n^c=0,
\end{equation}
which is satisfied due to the condition $n^a n_a=1$. Variation $\delta \phi^a$ gives us $n^a e_a|_\mathcal{Q}=0$, meaning that $n^a$ is orthogonal to the EOW brane $\mathcal{Q}$. Variation $\delta e$ results in a nontrivial constraint 
\begin{equation}
\varepsilon_{ab}(T n^a+\phi^a)-\varepsilon_{ac}\phi^a n^c n_b=0.
\end{equation}
Note that if we multiply the previous equation with $n_b$ we consistently get $0=0$.
Variation with respect to $\varphi$ results in the constraint (see Appendix B) 
\begin{equation}
\nabla_\mu n^\mu=0,
\end{equation}
meaning that the EOW is actually a geodesic. This coincides with the JT result. Moreover, this result was recently used in the analysis of the BF formulation of two-dimensional supergravity \cite{Belaey:2023jtr}, and we believe that our work provides necessary supporting arguments for the validity of that claim. The last variation $\delta n$ yields 
\begin{align}\nonumber&2\kappa\int_{\mathcal{Q}}\varepsilon_{ab}\varphi \delta n^a \Diff n^b+2\kappa \int_{\mathcal{Q}}\varepsilon_{ab}\varphi \Diff n^b\delta n^a\\\nonumber &-2\kappa \int_{\mathcal{Q}}\varepsilon_{ab}e^c\phi_c n^a\delta n^b+2\kappa T \int_{\mathcal{Q}} \varepsilon_{ab}\delta n^a e^b\\& - 2\kappa \int_{\mathcal{Q}}\varepsilon_{ab}\phi^a n^b\delta n_c e^c.
\end{align}
Using the identity 
\begin{equation}
    \varepsilon_{ab}\phi^c e_c n^a+\varepsilon_{ac}\phi^a n^ce_b=\varepsilon_{ab}\phi^b e_cn^c=0,
\end{equation}
since $e_cn^c\vert_{\mathcal{Q}}=0$,
we see that the variation is given by 
\begin{align}\nonumber
    &2\kappa \int_{\mathcal{Q}}\varepsilon_{ab}\varphi \delta n^a \Diff n^b+2\kappa \int_{\mathcal{Q}}\varepsilon_{ab}\varphi \Diff n^b\delta n^a\\&+2\kappa T \int_{\mathcal{Q}} \varepsilon_{ab}\delta n^a e^b.\label{pon}
\end{align}
Following \cite{Takayanagi:2020njm}, we should add a projector multiplying the equation that follows from 
this variation. This is due to the fact that $\delta(n_an^a)=0$. One might object by noting that we do not have to assume that $n^a$ is normalized, as the normalization follows from the equations of motion. However, when we wrote the boundary term, we had already assumed that the $n^a$ vector is normalised. Otherwise, we would have to properly normalise all the terms (see \cite{Wang:2016vid} for example). Either way we obtain the same equation,
\begin{equation}\label{identity}
2\varepsilon_{ab}\varphi\Diff n^b+T\left( \varepsilon_{ab}e^b-n^cn_a\varepsilon_{cb}e^b  \right)=0.   
\end{equation}
However, this equation is automatically satisfied given that $n_an^a=1$ and $n^ae_a|_{\mathcal{Q}}=0$, see Appendix B. Therefore, there are no new constraints. 

Finally, let us also demonstrate that by using the methodology from \cite{Erdmenger:2022nhz}, we obtain the same boundary term. First, due to general consideration in \cite{Erdmenger:2022nhz}, only $\int\varepsilon_{ab}\varphi R^{ab}$ term is relevant for computing the GHY boundary term. Assuming the notation from \cite{Erdmenger:2022nhz}, we have 
\begin{align}
\star \varphi_{\mathbf{n}a}\delta \varrho^{\mathbf{n}a}\sim \kappa \varphi\varepsilon_{\mathbf{n}a}\delta \varrho^{\mathbf{n}a}, \end{align}
which gives us 
\begin{equation}
\star \varphi_{\mathbf{n}a}=\kappa \varphi \varepsilon_{\mathbf{n}a},
\end{equation}
and the boundary term that we should add, from this point of view, is 
\begin{equation}
2\kappa \int_{\partial\mathcal{Q}}\varphi K^a \varepsilon_{\mathbf{n}a}=2\kappa \int_{\partial\mathcal{Q}}\varphi n^\mu D n^\nu \varepsilon_{\mu\nu},
\end{equation}
where we used the identity 
\begin{equation}
e^a_\nu Dn^\nu n^\mu \varepsilon_{\mu\rho}e^\rho_a=\varphi n^\mu D n^\nu \varepsilon_{\mu\nu}.    
\end{equation}
Therefore, all three viewpoints agree that (\ref{prvibound}) is the adequate boundary term for 2D BF gravity, supporting the claim that the EOW brane assumes the shape of a geodesic.

\section{Three-dimensional gravity with torsion}\label{sectorsion}

As a continuation of the work in \cite{Takayanagi:2020njm}, we will show that the AdS/BCFT construction can also be applied to bulk geometries with torsion. For that matter, consider the case of $\textrm{3D}$ AdS Chern-Simons (CS) gravity 
(which coincides with $\textrm{3D}$ EH gravity with a negative cosmological constant) 
modified by the translational CS term that involves torsion $T^{a}$. The bulk ($\mathcal{N}$) action is 
\begin{equation}\label{MB}
\kappa\int_{\mathcal{N}}\varepsilon_{abc}\left(R^{ab}e^c +\frac{1}{3}e^ae^be^c\right)+\alpha\kappa\int_{\mathcal{N}} T^ae_a,
\end{equation}
where $\alpha$ is a parameter of the theory. The translational CS term is part of the more general MB model of three-dimensional gravity with torsion \cite{Mielke:1991nn}, which has been studied from a holographic perspective in \cite{Klemm:2007yu}. 
By comparing action (\ref{MB}) and the MB model, it is clear that we are not including the $L_{\textrm{CS}}(\omega)$ term (the CS term involving the spin-connection as a gauge field). The reason for this is practical: this term depends explicitly on $\omega^{ab}$, not through the field strengths (curvature and torsion), and therefore the methodology of \cite{Erdmenger:2022nhz} is not applicable. Furthermore, it is obvious that for this term, one cannot move the variation at the boundary from $\omega$ to $e$, as this term involves only the $\omega$ field. Yet holography points out that a certain boundary term should be added. Namely, it follows from the considerations in \cite{Klemm:2007yu} that the $L_{\textrm{CS}}(\omega)$ term requires an additional boundary term of the form $\int \tilde{k}^i \tilde{e}_i$ 
and this term is not zero, even on-shell. 

Due to the translational CS term, the on-shell value of torsion will not be zero as for the pure $\textrm{3D}$ AdS CS gravity. Equations of motion obtained by varying (\ref{MB}) with respect to $e^{a}$ and $\omega^{ab}$ are given by 
\begin{align}\label{Reom}
&R^{ab}+\left( 1-\alpha^2  \right)e^ae^b=0, \\ \label{Teom}
&T^c=-\frac{\alpha}{2}\varepsilon^{abc}e_ae_b.
\end{align}
We see that torsion is not zero, but it is determined by the $e^{a}$ field, that is, by the metric structure of the theory. For $\alpha \rightarrow 0$, we obtain zero torsion as expected. Following \cite{Takayanagi:2020njm}, let us choose the bulk spacetime metric
\begin{equation}\label{metrika}
ds^{2}=e^{2\rho}(-\diff t^{2}+\diff\phi^{2})+l^{2}\diff\rho^{2},    
\end{equation}
for which we have 
\begin{equation}
e^0=e^\rho \diff t, \hspace{5mm} e^1=l\diff \rho, \hspace{5mm} e^2=e^\rho \diff \phi,
\end{equation}
where we introduced the length scale $l$. For pure AdS CS gravity (for $\alpha\rightarrow 0$), $l$ would be equal to $1$ (actually, to the length scale from the action that we conveniently set equal to $1$), and this would be the AdS radius. However, pure AdS$_{3}$ spacetime is not a solution to the theory (\ref{MB}).
Namely, although the metric is the same, the spin connection is modified due to the translational CS term,
\begin{align}\nonumber
&\omega^{01}=e^\rho\left(\frac{1}{l}\diff t -\frac{\alpha}{2}\diff \phi  \right), \\
&\omega^{02}=\frac{l\alpha}{2}\diff \rho, \\
& \omega^{12}=e^\rho\left( -\frac{\alpha}{2}\diff t -\frac{1}{l}\diff \phi \right).
\end{align}
The corresponding curvature is given by
\begin{equation}
R^{ab}=\left(\frac{\alpha^{2}}{4}-\frac{1}{l^{2}}\right)e^{a}e^{b}.    
\end{equation}
By comparing the equation of motion (\ref{Reom}) we see that the length scale $l$ from the metric depends on $\alpha$ as
\begin{equation}
\frac{1}{l^{2}}=1-\frac{3\alpha^{2}}{4}.    
\end{equation}
Now we come to the issue of boundary terms. As before, the boundary $\partial\mathcal{N}$ consists of two parts $\partial\mathcal{N}=\mathcal{M}\cup\mathcal{Q}$. Since we are interested in the profile of the EOW brane $\mathcal{Q}$, we will focus only on that part of the boundary. On the asymptotic boundary $\mathcal{M}$, we impose Dirichlet boundary conditions, as usual.  

\begin{figure}[h!]
    \centering
\includegraphics[width=0.45\textwidth]{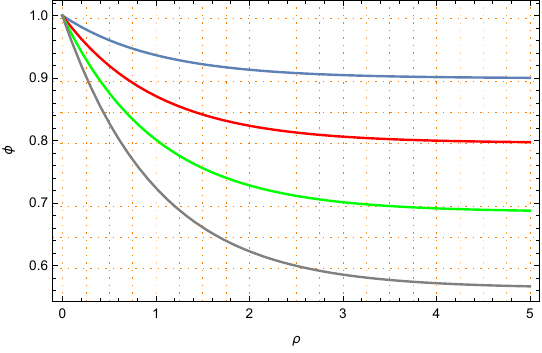}
    \caption{EOW brane profile for different values of parameter $\alpha$ (and, therefore, of $l$; here shown $l=1,2,3,4$ from top to bottom). For this plot, we set $T=0.1$, and fix the boundary of the CFT at $\phi=1$. We put all the brane profiles on the same graph even though they are defined on different spacetimes, as we can connect coordinate $\rho$ in those different spacetimes using the damping factor $e^\rho$ from the metric. }
    \label{profbrane}
\end{figure}

In the variation of action (\ref{MB}), there are two boundary terms,
\begin{equation}\label{bts}
-\kappa\int_{\mathcal{Q}}\varepsilon_{abc}e^{a}\delta\omega^{bc}-\alpha\kappa\int_{\mathcal{Q}}e_{a}\delta e^{a}.    
\end{equation}
If we pertain to the criteria that we should only have  $\delta e$ variation, we have to add a boundary term to action (\ref{MB}) in order to move the variation from $\omega$ to $e$ in the first boundary term from (\ref{bts}). It is easy to see that the appropriate boundary term is \cite{Corichi:2015nea} 
\begin{equation}
\kappa\int_{\mathcal{Q}}\varepsilon_{abc}e^{a}\omega^{bc}.\end{equation}
We already showed in (\ref{2C}) that this term follows from holography in the torsion-free case, but computation in \cite{Klemm:2007yu} shows that there should be no other boundary term even when $\alpha\neq 0$.
This seemingly non-covariant term can be cast in a covariant form using the normalized vector $n^{a}$ at the brane,
\begin{align}\label{bt}
&-2\kappa\int_{\mathcal{Q}}\varepsilon_{abc}e^{a}n^{b}\Diff_{\omega} n^{c}\\
&=-2\kappa\int_{\mathcal{Q}}\epsilon_{abc}e^{a}n^{b}\diff n^{c}-2\kappa\int_{\mathcal{Q}}\epsilon_{abc}e^{a}n^{b}\omega^{cd} n_{d}\nonumber\\
&=-2\kappa\int_{\mathcal{Q}}\epsilon_{abc}e^{a}n^{b}\diff n^{c}+\kappa\int_{\mathcal{Q}}e^a(\delta^b_a -n_an^b)\varepsilon_{bcd}\omega^{cd}.\nonumber
\end{align}
The last line can be obtained by introducing $\omega_{a}=\frac{1}{2}\varepsilon_{abc}\omega^{bc}$  or $\omega^{ab}=-\varepsilon^{abc}\omega_{c}$ \cite{Corichi:2015nea}. Namely, we have
\begin{align}
-2\varepsilon_{abc}e^{a}n^{b}\omega^{cd}n_{d}&=2\varepsilon_{abc}e^{a}n^{b}\varepsilon^{cdm}\omega_{m}n_{d}\nonumber\\
&=-2e^{a}n^{b}(\omega_{b}n_{a}-\omega_{a}n_{b})\nonumber\\
&=2e^{a}(n^{2}\delta^{b}_{a}-n_{a}n^{b})\omega_{b}\nonumber\\
&=e^{a}(\delta^{b}_{a}-n_{a}n^{b})\varepsilon_{bcd}\omega^{cd},
\end{align}
where we used the fact that $n^{2}=n_{a}n^{a}=1$. The expression $P^{b}_{a}=\delta^{b}_{a}-n_{a}n^{b}$  that appears in the final form of the boundary term (\ref{bt}) is the projector to the brane. The obtained term is precisely the GHY boundary term that follows from computations along the lines of \cite{Erdmenger:2022nhz} (again, the term that contains torsion does not lead to an additional boundary term). If we also include the tension of the brane, the total action is therefore
\begin{align}\nonumber
&\kappa\int_{\mathcal{N}}\varepsilon_{abc}\left(R^{ab}e^c +\frac{1}{3}e^ae^be^c\right)+\alpha\kappa\int_{\mathcal{N}} T^ae_a\\ &\nonumber-2\kappa\int_{\mathcal{Q}}\varepsilon_{abc}e^an^b\diff n^c+\kappa\int_{\mathcal{Q}}e^a(\delta^b_a -n_a n^b)\varepsilon_{bcd}\omega^{cd}\\ &+\kappa T\int_{\mathcal{Q}}\varepsilon_{abc}n^a e^be^c.
\end{align}
Variation $\delta \omega$ yields the constraint 
\begin{equation}
e_an^a|_{\mathcal{Q}}=0,
\end{equation}
which means that $n^a=n^\mu e_\mu^a$ is orthogonal to the EOW brane $\mathcal{Q}$. It is important to note that orthogonality is obtained dynamically; it has not been assumed ab initio. Variation $\delta n^a$ gives no further restrictions (it is identically satisfied given that $n^{2}=1$). Finally, the most important equation is obtained by variation $\delta e$,
\begin{equation}\label{eome}
P^b_a\varepsilon_{bcd}\omega^{cd}-2\varepsilon_{abc}n^b\diff n^c+2T\varepsilon_{abc}n^b e^c+\alpha e_a=0.
\end{equation}
To find the profile of the EOW brane, we assume that $\phi=g(\rho)$ and so $(\diff\phi-g'(\rho)\diff\rho)|_{\mathcal{Q}}=0$. The normalized vector orthogonal to the brane is therefore 
\begin{align}
n^{0}&=0,\nonumber\\
n^{1}&=-\frac{g'(\rho)}{l\sqrt{\frac{(g'(\rho))^{2}}{l^{2}}+e^{-2\rho}}},\nonumber\\
n^{2}&=\frac{e^{-\rho}}{\sqrt{\frac{(g'(\rho))^{2}}{l^{2}}+e^{-2\rho}}}. 
\end{align}
Consider now the equation (\ref{eome}) and take $a=2$. Using the fact that 
\begin{align}
n^{a}\omega_{a}&=-\frac{\alpha}{2}n^{a}e_{a}-\frac{1}{l}n^{0}e^{2}+\frac{1}{l}n^{2}e^{0}\nonumber\\
&=-\frac{1}{l}n^{2}e^{0},    
\end{align}
where the last step follows from $n^{0}=0$ and the constraint $n^{a}e_{a}\vert_{\mathcal{Q}}=0$, we get 
\begin{align}
g(\rho)=\pm\frac{l^{2}T}{\sqrt{1-l^{2}T^{2}}}e^{-\rho}+const.   
\end{align}
The equation (\ref{eome}) for $a=1$ further determines that the correct sign is plus. This gives us $n^{1}=lT$ and $n^{2}=\sqrt{1-l^{2}T^{2}}$. Finally, the equation for $a=0$ is now automatically satisfied. Therefore, the profile of the EOW brane is given by
\begin{equation}
\phi=\frac{l^{2}T}{\sqrt{1-l^{2}T^{2}}}e^{-\rho}+const,
\end{equation}
where $\frac{1}{l^{2}}=1-\frac{3\alpha^{2}}{4}$. We conclude that the modified affine structure (non-zero torsion) of AdS$_{3}$ spacetime leaves its mark in the EOW brane profile by changing the parameter $l$, which would otherwise be equal to $1$. There is a whole family of EOW branes parametrized by $\alpha$ from the translational CS term (Fig. \ref{profbrane}).  

\section{Summary and conclusions}

In this paper, we have revised the role of boundary terms in first-order gravity. We primarily focused on those cases where different criteria lead to the same conclusion, thus supporting the choice of the boundary term. Actually, in the context of AdS/BCFT duality, it is not clear how to choose boundary terms to get viable brane dynamics \cite{Hu:2022ymx}, and therefore the task is expected to be even harder in the first-order formulation. One way to see if the boundary term is appropriate, apart from being able to obtain a nontrivial EOW brane profile, is to compute the boundary stress-energy tensor and check if the BCFT boundary conditions are satisfied \cite{Izumi:2022opi}.

Based on this general discussion, we showed that by including the adequate boundary term, the profile of the EOW brane in the BF formulation of JT gravity indeed matches the geodesic, as expected. Using the same logic, we considered the modified version of the three-dimensional CS gravity with the translational CS term that explicitly involves torsion. The resulting theory has an AdS-like solution that exhibits non-zero torsion. We calculated the profile of the EOW brane for this geometry and found a family of solutions parametrized by the coupling constant $\alpha$ of the translational CS term from (\ref{MB}).
This extends the AdS/BCFT construction from \cite{Takayanagi:2020njm} to include bulk geometry with torsion. Note that the computations in this paper are consistent with the BCFT boundary conditions \cite{Izumi:2022opi}. The holographic stress-energy tensor was computed in \cite{Klemm:2007yu}. It is easy to verify that the one-point correlation function for the stress-energy tensor for the solution we analysed is zero and, therefore, trivially satisfies the BCFT boundary conditions.   

Further considerations could involve AdS/BCFT construction for more complicated systems, such as $\textrm{5D}$ CS gravity. It is not entirely clear to what extent the discussion regarding boundary terms made here applies in that particular case. Note that, even in the case of Riemannian gravity, it is not always clear which GHY-like term to add or what are the proper boundary conditions for bulk fields in order to obtain the EOW profile \cite{Hu:2022ymx, Miao:2017gyt}. Also, it is possible that, for $\textrm{5D}$ CS gravity, one will have to go beyond the idea, introduced in \cite{Takayanagi:2020njm}, to treat $n^a$ independently from the metric field. This paper illustrates that there are cases where one can safely apply this idea, but it is clear that it does not always hold. Moreover, we focused only on the spacetime (\ref{metrika}), but one should be able to discuss more general excited states. In the case at hand, this looks like a straightforward generalization.
Nevertheless, the rich structure of $\textrm{3D}$ gravity makes it an interesting and important model of first-order gravity that could reveal some new insights and lead us to a deeper understanding of the relevance of boundary terms.

\section*{Acknowledgements}

We thank Ioannis Papadimitriou, Bastian He{\ss} and Milutin Blagojevi\'{c} for useful discussions. Work of D.D. and D.G. is supported by the funding provided by the Faculty of Physics, University of Belgrade, through grant number 451-03-47/2023-01/200162 by the Ministry of Science, Technological Development and Innovations of the Republic of Serbia. The research was supported by the Science Fund of the Republic of Serbia, grant number TF C1389-YF, Towards a Holographic Description of Noncommutative Spacetime: Insights from Chern-Simons Gravity, Black Holes and Quantum Information Theory - HINT.  

\bibliographystyle{elsarticle-num} 
\bibliography{example}

\appendix
\setcounter{figure}{0}
\renewcommand{\thefigure}{A\arabic{figure}}
\renewcommand{\theequation}{\Alph{section}.\arabic{equation}}

\section*{APPENDIX}

\section{Algebraic structure of 2D BF theory}\label{A}

To begin with, let us consider the action for 3D Lovelock-Chern-Simons gravity with the negative cosmological constant,
\begin{equation}\label{LCS}
\int\varepsilon_{ABC}\left(R^{AB}e^{C}+\frac{1}{3}e^{A}e^{B}e^{C}\right).    
\end{equation}
The Lagrangian is invariant (up to a locally closed form) under $SO(2,2)$ gauge transformations (isometries of AdS$_3$ embedded in flat spacetime with signature $(-++-)$), which follows from the fact that fields $\omega^{AB}$ and $e^{A}$ can be regarded as components of an enlarged $SO(2,2)$ connection
\begin{equation}
\mathcal{A}=\frac{1}{2}\omega^{AB}J_{AB}+e^{A}J_{A3}, \end{equation}
where $J_{AB}$ and $J_{A3}$ are the group generators; indices $A,B$ take values $0,1,2$. Using the connection $\mathcal{A}$, we can construct the action for 3D Chern-Simons gauge theory, which is the same as (\ref{LCS}) up to a boundary term.   

By performing Kaluza-Klein dimensional reduction (together with truncation of certain fields) of (\ref{LCS}) one obtains 2D BF action 
\begin{align}\label{2DBF}
\int \varepsilon_{\hat{A}\hat{B}\hat{C}}\phi^{\hat{A}}F^{\hat{B}\hat{C}},  
\end{align}
invariant under $SO(1,2)$ gauge transformations generated by $J_{\hat{A}\hat{B}}=(J_{ab},J_{a3})$ with Lorentz $SO(1,1)$ indices $a,b=0,1$. Note that $SO(1,2)$ indices (denoted with the hat symbol) take values $\hat{A},\hat{B}=0,1,3$, and we define $\varepsilon_{013}=1$.

The $SO(1,2)$ field strength is given by
\begin{align}
F=F^{\hat{a}\hat{B}}J_{\hat{A}\hat{B}}=\frac{1}{2}\left(R^{ab}+e^{a}e^{b}\right)J_{ab}+T^{a}J_{a3},
\end{align}
with curvature and torsion,
\begin{align}
R^{ab}&=\diff\omega^{ab}+\omega^{a}_{\;\;c}\omega^{cb},\\
T^{a}&=\Diff e^{a}=\diff e^{a}+\omega^{a}_{\;\;b}e^{b},
\end{align}
while the multiplet of (spacetime) scalar fields is organized as  
\begin{align}
\Phi=\phi^{\hat{A}}J_{2\hat{A}}=\phi^{a}J_{2a}+\varphi J_{23}.
\end{align}
Action (\ref{2DBF}) can thus be written in a manifestly $SO(1,2)$ invariant form,
\begin{equation}
\int\tr\left(\Phi F\right).   
\end{equation}
In terms of Lorentz tensors, 2D BF action can be formulated as (\ref{bf}).

\section{GHY term in 2D gravity}\label{B}

Here, we explicitly prove the identification of the GHY term in the metric formulation of gravity with the expression $2\kappa \int \varphi\varepsilon_{ab}n^a\Diff n^b$. We have 
\begin{align}\nonumber
2\kappa \int_\mathcal{Q} \varepsilon_{ab}\varphi n^a\Diff n^b&=
2\kappa \int_\mathcal{Q} \varepsilon_{ab}\varphi e^a_\rho n^\rho\Diff_\mu(e^b_\nu n^\nu) \diff x^\mu\\\nonumber
&=2\kappa \int_\mathcal{Q} \varepsilon_{ab}\varphi e^a_\rho n^\rho e^b_\nu \nabla_\mu n^\nu \diff x^\mu\nonumber\\\nonumber
&=2\kappa \int_\mathcal{Q} e\varepsilon_{\rho\nu}\varphi n^\rho  \nabla_\mu n^\nu \diff x^\mu\\
&=2\kappa\int_\mathcal{Q} e\varepsilon_{\rho\nu}\varphi n^\rho  \nabla_\mu n^\nu \diff \frac{\diff x^\mu}{\diff t}\diff t.
\label{b1}
\end{align}
With $\Diff_\mu$ we denote the Lorentz covariant derivative, with $\nabla_\mu$ the spacetime covariant derivative with connection $\Gamma_{\mu\nu}^\lambda$, and $t$ parameterizes one-dimensional boundary.
Now we use the fact that metric $g_{\mu\nu}$ can be written as $g_{\mu\nu}=h_{\mu\nu}+n_\mu n_\nu$, where we can further write $h_{\mu\nu}=-t_\mu t_\nu$, with $t_\mu$ unit tangent vector to the one-dimensional boundary \cite{Gao:2021uro} (not insisting on a normalisation of vector $t^\mu$ would introduce the factor of $\sqrt{|h|}$ in the boundary term). Note that we have 
\begin{equation}
n^ae_a|_{\mathcal{Q}}=g_{\mu\nu}n^\mu \diff x^\nu |_{\mathcal{Q}}=0, 
\end{equation}
from which we conclude $n^\mu \perp \frac{\diff x^\mu}{\diff t}\equiv t^\mu$. Next, we define epsilon tensor $\overline{\varepsilon}_{\mu\nu}=e\cdot \varepsilon_{\mu\nu}$ and prove that
\begin{equation}\nonumber
\overline{\varepsilon}_{\mu\nu}n^\mu=-t_\nu.
\end{equation}
First, note that if we multiply both sides by $n^\nu$, we consistently get $0=0$. Therefore, $\overline{\varepsilon}_{\mu\nu}n^\mu$ is proportional to $t_{\nu}$. We furthermore fix the normalisation by computing 
\begin{align}\nonumber 
&\overline{\varepsilon}_{\mu\nu}n^\mu\overline{\varepsilon}_{\rho\sigma}n^\rho g^{\nu\sigma}=-\overline{\varepsilon}_{\mu\nu}\overline{\varepsilon}^{\nu}_{\;\;\rho}n^\mu n^\rho\\\nonumber 
&=g_{\mu\rho}n^\mu n^\rho=1.
\end{align}
Therefore, the constant of proportionality is one (or minus one, which is just a different choice of orientation). This implies that (\ref{b1}) can be written as 
\begin{align}
    &2\kappa \int_\mathcal{Q} \varphi t_{\nu}  \nabla_\mu n^\nu \diff x^\mu \nonumber\\
    &= 2\kappa \int_\mathcal{Q} \varphi t_{\nu} t^\mu \nabla_\mu n^\nu \diff t=2\kappa \int_\mathcal{Q}  \varphi K,
\end{align}
where $K=\nabla_\mu n^\mu$. This coincides with the usual GHY boundary term in the metric formulation and proves that the variation $\delta \varphi$ leads to the condition $\nabla_\mu n^\mu=0$.

Next, we prove that variation $\delta n$ does not lead to any new constraints. 
First, we prove that $\varepsilon_{ab}e^b-n^cn_a\varepsilon_{cb}e^b=0$ by explicitly writing down all indices, which can be easily done as we work in a low number of dimensions. For example, for $a=0$ we have
\begin{align}\nonumber
    &\varepsilon_{0b}e^b-n^cn_0\varepsilon_{cb}e^b\\\nonumber&=e^1-n^0n_0\varepsilon_{01}e^1-n^1n_0\varepsilon_{10}e^0\\\nonumber &=e^1-(1-n^1n_1)e^1+n^1n_0e^0\\ &=e^1-(1-n^1n_1)e^1-n^1n_1e^1=0.
\end{align}
In this derivation, we used the fact that $n_an^a=1$, and $n^ae_a|_{\mathcal{Q}}=0$. The same holds for $a=1$. Therefore,  
\begin{equation}\label{result1}
\varepsilon_{ab}e^b-n^cn_a\varepsilon_{cb}e^b=0.   \end{equation}
To prove that $\Diff n^a=0$ follows from the fact that $\varepsilon_{ab}n^a\Diff n^b=0$, we can again write down all the indices explicitly, 
 \begin{align}\nonumber
\varepsilon_{ab}n^a\Diff n^b&=n^0\diff n^1-n^1\diff n^0-\omega_{10}(n^0n_0+n^1n_1)\\ &=n^0\diff n^1-n^1\diff n^0-\omega_{10}=0.\label{b3}
 \end{align}
Multiplying this equation by $n_0$ we get
\begin{align}\nonumber 
    &n^0n_0\diff n^1-n^1n_0\diff n^0-\omega_{10}n_0\\\nonumber 
    =\;&(1-n^1n_1)\diff n^1-n^1n_0\diff n^0+\omega^{1}_{\;\;0}n^0\\\nonumber 
    =\;&\diff n^1 -\frac{1}{2}n^1\diff(n^an_a)+\omega^{1}_{\;\;0}n^0\\
    =\;&\diff n^1 +\omega^1_{\;\;0} n^0=\Diff n^1=0.
    \end{align}

Similarly, if we multiply (\ref{b3}) by $n^1$, we get 
\begin{align}
    &-\diff n^0+\frac{1}{2}n^0(n_an^a)-\omega_{10}n^1=-\Diff n^0=0.
\end{align}
This proves that $\Diff n^a=0$ and, together with (\ref{result1}), implies the identity (\ref{identity}).   

\end{document}